\documentstyle[12pt]{article}
\let\section=\subsection  \let\subsection=\subsubsection

\begin{document}
\begin{center}
  {\large  \bf {NEUTRON STAR PROPERTIES WITH NUCLEAR}}\\[2mm] 
  {\large  \bf {BHF EQUATIONS OF STATE}}\\[5mm]
  {M.~Baldo$^a$, G.~F.~Burgio$^a$, H.~Q.~Song$^b$ and F.~Weber$^c$}\\[5mm]
{$^a$ \small \it I.N.F.N. Sezione di Catania, c.so Italia 57, 
I-95129 Catania, Italy \\[2mm]}

{$^b$ \small \it Institute of Nuclear Research, Academia Sinica,  
Shangai 201800, China \\[2mm]}

{$^c$ \small \it Sektion Physik, Universit\"at M\"unchen, 
Am Coulombwall 1, D-85748 Garching, Germany \\[8mm]}
\end{center}

\begin{abstract}\noindent
We study the properties of static and rotating neutron stars adopting 
non-relativistic 
equations of state (EOS) for asymmetric nuclear matter based on the
Brueckner-Hartree-Fock (BHF) scheme. The BHF calculation, with the
continuous choice for the single particle potential, appears to be very
close to the full EOS, which includes the three-hole line contribution
calculated by solving the Bethe-Fadeev equations within the 
gap choice for the single particle potential.
Three-body forces are included in order to reproduce the correct 
saturation point for nuclear matter. 
A comparison with fully relativistic many-body calculations
of nuclear matter EOS is made.
\end{abstract}

\section{Introduction}
The properties of neutron stars such as masses and radii depend on the 
equation of state (EOS) at densities up to an order of magnitude higher 
than those observed in ordinary nuclei. Therefore the knowledge of the
EOS in the superdense regime is fundamental for the study of 
astrophysical compact objects \cite{sha}. 
For this purpose we derive an equation of
state for asymmetric nuclear matter, using a non-relativistic many-body 
theory within the framework of the Brueckner-Hartree-Fock (BHF) scheme
\cite{bethe,Baldo91}.
In this approach, the basic input is the two-body nucleon-nucleon (NN)
interaction. The BHF approximation, with the continuous choice for the 
single particle potential, reproduces closely the many-body calculations up to
three hole-line level \cite{song}.
However, as it is well known, the empirical saturation
point is not reproduced. Therefore we have included a contribution 
coming from three-body forces to reproduce the correct saturation point
\cite{BBB}.
Those EOS's are the fundamental input for constructing models of 
static and rotating neutron stars in the framework of Einstein's 
theory of general relativity by applying a refined version of Hartle's stellar
structure equations \cite{wg,har}.
We calculate properties of neutron stars like 
gravitational mass, equatorial and polar radius, 
for sequences of star models either static or rotating at their
respective general relativistic Kepler frequencies.
We compare with predictions from fully relativistic
microscopic EOS.
\noindent

\section{Equation of state}

Microscopic calculations of nuclear matter EOS have been performed in the
framework of the Brueckner--Hartree--Fock (BHF) scheme \cite{bethe}. 
The energy per particle $E/A$ within the BHF scheme is given
in terms of the so-called reaction matrix G. The latter is obtained
by solving the Brueckner--Bethe--Goldstone (BBG) equation 

\begin{equation}
G(\omega) = V  + V \frac{Q}{\omega - H_0} G(\omega), \label{eq:gmat} 
\end{equation}                                                           
\noindent
where $\omega$ is the unperturbed energy of the interacting nucleons,
$V$ is the free nucleon-nucleon (NN) interaction, 
$H_0$ is the unperturbed energy of the intermediate scattering states
and Q is the Pauli operator which prevents scattering into occupied 
states. With the $G$-matrix we can calculate the total energy per nucleon
 
\begin{equation}
\frac{E}{A} =  \frac{3}{5} \frac{\hbar^2 k_F^2}{2m}  + U(n),\label{eq:ebind}
\end{equation}
\noindent
being $U(n)$ the contribution of the potential energy to the total energy
per particle

\begin{equation}
U(n) = {{1}\over{2A}} \sum_{k,k'\leq k_F} \langle k k'|G(\omega=\epsilon(k)
+\epsilon(k'))|k k'\rangle_a \label{eq:ubind}
\end{equation}
\noindent
where the subscript {\it a} indicates antisymmetrization of the 
matrix element.
The single-particle energies are denoted by $\epsilon$.
In this scheme, the only input quantity we need is the bare NN interaction
$V$ in the Bethe-Goldstone equation (1). 

The Brueckner-Hartree-Fock (BHF) approximation for the EOS in symmetric 
nuclear matter, within the 
continuous choice \cite{Baldo91}, reproduces closely results which 
include up to three hole-line diagram contributions to the BBG expansion of 
the energy, calculated within the so called {\it gap choice} for the single 
particle potential \cite{song}. 
In Fig.1 we show the energy per nucleon calculated within this scheme 
in the case of symmetric and neutron matter  
using the Argonne $v_{14}$ model \cite{Wir84} for the two-body 
nuclear force. The open squares represent the solution 
of the Bethe-Fadeev equations for the three hole-line in the gap 
choice. These results extend to higher densities the calculations 
published in ref.\cite{song} for symmetric nuclear matter.
Recently it has been shown that the results up to three hole-lines
are independent from the choice of the single particle potential \cite{prl},
which gives evidence of convergence of the BBG expansion.

For neutron matter the results are preliminary. In any case we found a
negligible difference between the BHF results in the continuous and in the
gap choice. Correspondingly the three hole-line contribution in neutron
matter turns out to be much smaller than in symmetric nuclear matter.
All these results together give support to the use of the BHF
approximation in the continuous choice in the study of neutron stars.

We notice that the BHF fails
to reproduce the empirical saturation point of nuclear matter.
This well known deficiency, which does not depend on the choice of the
two-body force, is commonly corrected introducing 
three-body forces (TBF). We adopted the Urbana three-nucleon model
\cite{Car83}, which consists of an attractive term due to 
two--pion exchange with excitation of an intermediate $\Delta$-resonance, 
and a repulsive phenomenological central term.
Several details are given in ref.\cite{BBB}. 

The corresponding EOS obtained using
the $v_{14}$ potential is depicted in Fig.2a)
for symmetric and neutron matter (solid line).
This EOS saturates at  $n_o = 0.178~fm^{-3},~ E_o/A = -16.46~MeV$
and it is characterized by an incompressibility 
$K_{\infty} = 253~MeV $. 
In Fig.2a) we plot also the EOS from a recent Dirac-Brueckner calculation 
(DBHF) \cite{Li92} with the Bonn--A two--body force 
(dashed line).      
In the low density region the BHF equation 
of state with TBF and DBHF equation of state are very similar, 
whereas at higher density the DBHF is stiffer.  
The discrepancy between the non-relativistic and relativistic 
calculation of the EOS can be
easily understood by noticing that the DBHF treatment is equivalent
\cite{Baldo95} to introduce in the non-relativistic BHF the three-body
force corresponding to the excitation of a nucleon-antinucleon
pair, the so-called Z-diagram which is repulsive
at all densities. In BHF treatment, on the contrary, both attractive and 
repulsive three-body forces are introduced, and therefore a softer 
EOS is expected. 

The properties of neutron stars (NS) depend on the knowledge of the
EOS over a wide range of densities, {\it i.e.} from the density of iron
at the star's surface up to several times the density of normal nuclear
matter encountered in the core \cite{sha}.
It is commonly accepted that the interior part of a neutron star is made 
mainly by nuclear matter (eventually superfluid) with a certain lepton 
fraction, although the high-density core might suffer a transition to 
other hadronic components. Here we assume that a neutron star is composed 
only by nucleons and leptons, {\it i.e.} an uncharged mixture of neutrons,
protons, electrons and muons in $\beta$-equilibrium. The presence of
leptons softens the EOS with respect to the pure neutron matter case.
In Fig.2b) the EOS for $\beta$-stable matter is shown for the cases 
previously discussed. It has to be stressed that in $\beta$-stable
matter the density dependence of the nuclear symmetry energy affects the 
proton concentration \cite{bl}. As it has been recently pointed out by 
Lattimer et al. \cite{lat}, the value of 
the proton fraction in the core of NS is crucial for the onset of 
direct Urca processes, whose occurrence enhances neutron star cooling rates.

From the energy per baryon of asymmetric nuclear matter in 
$\beta$-equilibrium, we calculate the nuclear contribution $P_{nucl}$ to  
the total pressure of stellar matter as well as the mass density 
$\rho_{nucl}$.  
Then the total pressure and total mass density can be easily calculated 
by adding the leptonic contributions. For more details, see ref.\cite{BBB}.

\section{Neutron star structure }

Neutron stars are objects of highly compressed matter so that the geometry of
space-time is changed considerably from flat space. Thus Einstein's general 
theory of relativity must be applied. Therein the Einstein curvature
tensor ${\cal G}_{\mu\nu}$ is coupled to the energy-momentum density
tensor ${\cal T}_{\mu\nu}$ of matter (G denotes the gravitational constant):

\begin{equation} 
{\cal G}_{\mu\nu} = 8~\pi~G~{\cal T}_{\mu\nu}(\rho, P(\rho)) \label{eq:eins}
\end{equation}
\noindent
The knowledge of the EOS, {\it i.e.} pressure $P$ as function of the 
energy density $\rho$ is therefore required in order to solve 
eq.(\ref{eq:eins}). For a spherically symmetric and static star, Einstein's
equations reduce to the familiar Tolman--Oppenheimer--Volkoff (TOV)\cite{tov}
equations 

\begin{equation}
{dP(r)\over{dr}} = -{{G m(r) \rho(r)} \over{r^2}} 
{ {(1 + {P(r)\over {c^2 \rho(r)}})} 
{(1 + {4\pi r^3 P(r)\over {c^2 m(r)}})} \over
{ {(1 - {2G m(r)\over {r c^2 }})} }} \label{eq:tov1}
\end{equation}

\begin{equation}
{dm(r)\over{dr}} = 4 \pi r^2 \rho(r) \label{eq:tov2}
\end{equation}
\noindent

For a given EOS {\it i.e.} $P(\rho)$, one can solve the TOV equations by 
integrating them for a given central energy density $\rho$, from the star's
center to the star's radius defined by $P(R_s)=0$.
This gives the stellar radius $R$ and the gravitational mass 
is then 
\begin{equation}
M_G~ \equiv ~ m(R)  = 4\pi \int_0^Rdr~ r^2 \rho(r). 
\end{equation}
\noindent

The case of rotating stars is more complicated, since changes occur in the 
pressure, energy density and baryon number density because of the rotation. 
In this work we adopt the method developed in ref.\cite{wg}, which is a 
redefined version of Hartle's perturbative method \cite{har} for the
investigation of the general relativistic Kepler frequency of a rotating 
neutron star. We recall that for rotation at frequencies beyond the 
Kepler value $\Omega_K$, mass shedding at the equator sets in which makes 
the star unstable. Therefore $\Omega_K$ sets an absolute upper bound on the 
rotational frequency. For every nuclear EOS there are uniquely determined 
values of $\Omega_K$ for each star in the sequence up to the limiting 
mass value. 
In Fig.3 we show the results obtained with our BHF plus TBF
equation of state for asymmetric nuclear matter and with the Dirac-Brueckner
one.
We display the gravitational mass $M_G$, in units of the solar mass 
$M_{\odot}$ ($M_{\odot} = 1.99~10^{33}$ g),  as a function 
of the radius R and the central energy density $\rho$
(in units of $\rho_0~=~140~MeV/fm^3$).
The upper lying curves show the increase of mass due to rotation
at the (absolute limiting) Kepler frequency, {\it i.e.} $\Omega = \Omega_K$.
The solid (dashed) line indicates a sequence of star models obtained with the
BHF+TBF (DBHF) equation of state. We observe larger gravitational 
masses, {\it i.e.} up to 14\% mass increase for BHF+TBF and 15\% 
for DBHF, relative to the spherical (non-rotating) Oppenheimer-Volkoff star
model of the same $\rho$ value. The large mass increase obtained is 
accompanied by relatively large (equatorial) radius value.
Moreover one sees that, for a fixed value of the gravitational mass,
the central star density $\rho$ decreases for increasing values of the
rotational frequency. This is because of the centrifugal force acting on the
star's matter together with the nuclear force. Their intensity must be 
counterbalanced by the attractive gravitational forces.
More details on rotating stars will be given in a forthcoming paper\cite{pap}. 

\section{Conclusions}

In conclusion, we computed some properties of NS's on the basis of a 
microscopic EOS obtained in the framework of BHF many--body theory with two
plus three--body nuclear interactions. Our EOS with three-body forces is able 
to reproduce the correct saturation point of nuclear matter. 
The comparison with the DBHF method shows that the relativistic effects 
represent a particular repulsive three-body force and gives rise to a
stiffer EOS. Therefore the predicted values for the limiting mass and radius
of neutron stars are higher in the DBHF case than in the non-relativistic 
BHF calculation. The inclusion of rotation substantially changes the
limiting values of the gravitational mass by about 14-15 \% for 
configurations rotating at their respective general relativistic
Kepler frequency. A more detailed comparison between the predictions 
of non-relativistic
BHF microscopic EOS and observational data on pulsars, {\it i.e.} fast
rotation and large enough neutron star masses, is currently in progress.

\newpage
\centerline {\bf Figure captions}

\vskip 1cm

{Fig.1 : The energy per baryon E/A is plotted vs. the 
number density n  for symmetric matter (lower curve) and 
for neutron matter (upper curve). The solid line represents
a Brueckner calculation with Av14 potential with the continuous choice.
The squares are the solutions of the Bethe-Fadeev equations with the
gap choice.} \par\noindent

{Fig.2 : The energy per baryon E/A is plotted vs. 
the number density n  in panel (a) for symmetric matter (lower curves) and 
for neutron matter (upper curves). The solid line represents 
non-relativistic BHF calculations with three-body forces and the dashed line 
a relativistic Dirac-Brueckner one. Panel (b): as in panel (a) but 
for $\beta$-stable nuclear matter.}\par\noindent 

{Fig.3 : The gravitational mass $M_G$, expressed in units
of the solar mass $M_{\odot}$, is displayed vs. radius R (panel (a))
and the central energy density $\rho$ (in units of $\rho_0~=~140
MeV/fm^3$) (panel (b)) for sequences of star models constructed for  
BHF+TBF (solid line) and DBHF (dashed line) equations of state. 
The two (lower) upper lying curves refer to (non-) rotating star models.}

\end{document}